\documentclass[arguments]{aastex631}
\usepackage[version=4]{mhchem}
\usepackage{chemformula}
\usepackage{chngcntr}
\usepackage{amsthm, amssymb}
\newcommand{\R}{\mathbb{R}}
\usepackage{algorithm2e}
\usepackage{extarrows}

\newtheorem{Definition}{Definition}
\newtheorem{Theorem}{Theorem}
\newtheorem{Lemma}[Theorem]{Lemma}

\newtheorem{myexample}{Example}
\newtheorem*{myexample*}{Example}
\usepackage{kbordermatrix}

\shorttitle{Input-output Monotonicity}
\shortauthors{Nasti et al.}
\graphicspath{{./}{figures/}}

\begin{document}
\title{Efficient Analysis of Chemical Reaction Networks Dynamics based on Input-Output Monotonicity}

\correspondingauthor{Lucia Nasti}
\email{lucia.nasti@di.unipi.it}

\author[0000-0003-4687-024X]{Lucia Nasti}
\affiliation{Department of Computer Science - University of Pisa \\
Largo Bruno Pontecorvo, 3, \\
56127, Pisa, Italy}

\author[0000-0002-7424-9576]{Roberta Gori}
\affiliation{Department of Computer Science - University of Pisa \\
Largo Bruno Pontecorvo, 3, \\
56127, Pisa, Italy}
\author[0000-0002-7309-6424]{Paolo Milazzo}
\affiliation{Department of Computer Science - University of Pisa \\
Largo Bruno Pontecorvo, 3, \\
56127, Pisa, Italy}
\author[0000-0003-2291-990X]{Federico Poloni}
\affiliation{Department of Computer Science - University of Pisa \\
Largo Bruno Pontecorvo, 3, \\
56127, Pisa, Italy}



\begin{abstract}

\textbf{Motivation:} A Chemical Reaction Network (CRN) is a set of chemical reactions, which can be very complex and difficult to analyze. Indeed, dynamical properties of CRNs can be described by a set of non-linear differential equations that rarely can be solved in closed-form, but that can instead be used  to reason on the system dynamics. In this context, one of the possible approaches is to perform numerical simulations, which may require a high computational effort. In particular, in order to investigate  some  dynamical properties, such as  robustness or global sensitivity, many simulations have to be performed   by varying the initial concentration of chemical species.\\
 \textbf{ Results:} In order to reduce the computational effort required when many simulations are needed to assess a property, we  exploit a new notion of  monotonicity  of the \textit{output} of the system (the concentration of a target chemical species at the steady state) with respect to the \textit{input} (the initial concentration of another chemical species). To assess such  monotonicity behaviour,  we  propose a new graphical approach that allows us to state  sufficient conditions for ensuring  that the monotonicity property holds. Our sufficient conditions allow us  to efficiently  verify the monotonicity property    by exploring a graph constructed on the basis of the reactions involved in the network. Once established, our monotonicity property allows us to drastically reduce the number of simulations required to assess some dynamical properties of the CRN.
\end{abstract}



\section{Introduction}\label{sec:Introduction}
From the discovery of DNA structure, in 1953, there has been a growing interest in  understanding  the morphological and functional organization of living cells \citep{kitano2002systems}. Cells are very complex to analyze because they consist of many components that interact with each other, through multiple sequences of chemical reactions, Chemical Raction Networks (CRNs), which regulate the overall behavior. Besides, several fluctuations can alter the cell functionalities, such as internal errors propagation and variation in the concentrations of chemical species.


Bioinformatics and systems biology emerge as powerful tools to investigate CRN dynamics merging computational methods and real data. Through simulations, for instance, it is possible to mimic the internal dynamics of a natural system and, therefore,  to predict its behaviour. Moreover, model-based analysis techniques can be used to interpret some less intuitive features of the system.

In this context, computer scientists developed many formalisms to study systems of interacting components, which can be applied to model and  describe  CRNs and, in general, biological systems. Among these formalisms, those that have been applied in systems biology \citep{bernini2018process} include Petri nets \citep{behinaein2014structural,murata1989petri,koch2010petri}, Hybrid systems \citep{alur1993hybrid,henzinger2000theory, li2017review},  process calculi such as the $\pi$-calculus \citep{regev2000representation} as well as many ad-hoc biologically inspired calculi \citep{danos2008abstract} and rule-based systems \citep{barbuti2011spatial}.

The development of models that can help in predicting the system behaviour  requires precise and detailed  information about the set of initial conditions of the  biological system under study.  In many cases, obtaining such precise informations is  unfortunately very challenging (or even impossible) because of the noisy nature of biological data \citep{dresch2010thermodynamic}. Moreover, some parameters may be affected by fluctuations that alter ordinary system behavior. Finally, in many cases such precise  informations simply cannot be measured. 

In order to determine which model parameters are more critical in case of perturbations and approximate measurements, it is possible to apply \textit{sensitivity analysis} methods, which give a measure of the behavioral change of the system under perturbation of one of its parameters. As underlined in \citep{iooss2015review,zi2011sensitivity}, there exist two fundamental approaches to  sensitivity analysis, \textit{local} and \textit{global}. While the local sensitivity analysis  investigates  the effects of small  perturbations, the global one studies the effects of  large  perturbations and  determines also  the most (or the least) influencing  parameters.


Sensitivity analysis methods (in particular, the global ones) typically require performing many simulations, by varying the system parameters one by one. These methods are, in general, quite expensive, because of the number of parameters and the large range of values to be tested. 

For these reasons, alternative dynamical properties of CRNs, such as  monotonicity \citep{angeli2006structural, gori2019towards} and steady-state reachability \citep{feinberg1987chemical}, have been studied. Establishing such properties, indeed, provides information on the CRN dynamics without the need of performing several numerical simulations \citep{nastiTesi}.

Monotonicity, in particular, is a property stating that a given measurable aspect of the system dynamics increases (or decreases) with the increase of a given system parameter. Many more specific definitions of monotonicity exist \citep{angeli2006structural, gori2019towards} and in some cases they can be tested simply by inspecting the structure of the system models. It is worth noting that the validation of  many different biological properties, such as robustness \citep{kitano2002systems, rizk2011continuous, shinar2010structural}, persistence \citep{angeli2007petri}, and adaptation \citep{shinar2009sensitivity},  greatly benefits from the assessment of  monotonicity properties of the network, since this typically allows reducing the number of cases to be analyzed through numerical simulations.

Consider, for example, the robustness property. It is observed in many biological systems and it expresses the ability of the system to preserve its functions despite the presence of perturbations \citep{kitano2007towards}. 
Without any assumption on monotonicity, verifying robustness would require, in general, to consider all possible perturbations, usually expressed as different initial states of the system. In particular, regarding the CRNs, it would be necessary to test the system behaviour by examining all the possible combinations of initial concentrations of chemical species and, in practice, this would require a huge (in principle, infinite) number of simulations \citep{nasti2018formalizing}.

The same reasoning applies also to the case in which the  initial concentrations of a biological network are simply unknown.   Without any assumption on monotonicity, understanding  the qualitative  system behaviour would  require, in principle, to consider all possible initial concentrations. 

In this paper we propose a sufficient condition for CRNs that, if satisfied, ensures that a form of monotonicity, called \textit{Input-Output monotonicity}, holds. Given a species considered as the \emph{input} of a CRN, and another species considered as the \emph{output}, we say that input and output are in a monotonicity relation if the concentration of the output species always increases (or decreases) in response to an increase in the initial concentration of the input. If this monotonicity relationship holds, than it is possible to consider an interval of initial concentrations for the input species and obtain the corresponding concentrations interval for the output species by performing only two simulations, one for each extreme value of the input initial concentration \citep{gori2019towards}.

The sufficient condition we propose is based on a condition on the structure of the CRN that can be efficiently evaluated, without the need of performing any simulation. Following the lines of \citep{angeli2006structural}, our condition is based on a graph representation of the CRN enriched with information about cooperation and competition among reactions and it is  expressed as a set of constraints on the graph structure.

 

The paper is organized as follows. In the rest of this section we recall some background definitions of CRNs and related concepts.
In Section~\ref{sec:Results} we define Input-Output monotonicity and state Theorem~\ref{THEOREM} expressing our sufficient condition. In Section~\ref{sec:Case}, we apply our methodology to study the case of the ERK signaling pathway. In Section~\ref{sec:Disc_Concl} we draw our conclusions and discuss future work. Finally, in Section~\ref{sec:Methods}, we give the proof of Theorem~\ref{THEOREM}.

\subsection{Chemical Reaction Networks} \label{sec:CRN}
A Chemical Reaction Network (CRN) is a set of reactions that can be  formally  defined as follows.  
\begin{Definition}[Chemical reaction network]
Given an indexed set of chemical species $\mathcal{S} = (\mathcal{S}_1,\mathcal{S}_2,\dots,\mathcal{S}_s)$, a \emph{Chemical Reaction Network} is an indexed set of reactions $\mathcal{R} = (\mathcal{R}_1,\mathcal{R}_2,\dots, \mathcal{R}_{r})$ involving such species. Each chemical reaction is denoted as
\[
    \mathcal{R}_i : \sum_{j=1}^s \alpha_{ij} \mathcal{S}_{j} \rightarrow \sum_{j=1}^s\beta_{ij}\mathcal{S}_{j}
    \qquad \mbox{ or } \qquad
    \mathcal{R}_i : \sum_{j=1}^s \alpha_{ij} \mathcal{S}_{j} \rightleftharpoons \sum_{j=1}^s\beta_{ij}\mathcal{S}_{j},
\]
where $\alpha_{ij}$ and $\beta_{ij}$ are non-negative integers called the stoichiometric coefficients. 
\end{Definition}
\noindent The arrow is used to indicate the direction in which a chemical reaction takes place (from reactants to products). When we have a single arrow ($\rightarrow$), the reaction is \textit{irreversible}, namely that a reaction transforming products back into reactants cannot take place. When there is a double arrow ($\rightleftharpoons$) it means that it is possible to have both the forward and the backward transformation, and then the reaction is \textit{reversible}. In a CRN, a reversible reaction can be equivalently represented as a pair of irreversible reactions with opposite directions.

The \emph{stoichiometric coefficient matrix} $\Gamma \in \mathbb{R}^{s\times r}$ is defined as 
\[
\Gamma_{ji}= \beta_{ij}- \alpha_{ij}.
\]
Note that we allow in our networks the presence of \emph{promoters}: a promoter is a species $\mathcal{S}_j$ (such as an enzyme) that affects the rate of a reaction $\mathcal{R}_i$ but does not get produced nor consumed by it, i.e., $\Gamma_{ji}=0$.

The \emph{rate} of a reaction is expressed as a function of the concentrations of the species in the network. The vector of species concentrations at a given time is denoted by $S= (S_1, S_2,..., S_{s})' \in \mathbb{R}^s$. The vector of reaction rates (which are functions of $S$) is denoted by $R(S)= (R_1(S), R_2(S),..., R_{r}(S))' \in \mathbb{R}^r$.

Following~\citep{angeli2010graph}, we assume that all entries of the Jacobian of $R(S)$ (which we denote by $DR\in\mathbb{R}^{r\times s}$) have a well-defined constant sign that does not depend on $S$ (although they may become zero for certain values of $S$), and that
\begin{equation} \label{assumption}
    DR_{ij} \Gamma_{ji} \leq 0 \quad \text{for all $i,j$}.
\end{equation}
The assumption~\eqref{assumption} is a natural one because in most cases the rate of a reaction increases with the quantity of reactants. This assumption covers, in particular, the most common case, described by the well-established law of \emph{mass action}, in which the rate of a reaction is proportional to the concentration of each of its reactants.

Following a deterministic approach, the evolution of the concentrations in time is usually described by a system of differential equations:
\[
	\frac{dS}{dt} = \Gamma \cdot{R(S)}, \quad  S \in \R_{\geq 0}^{s}.
\]

\begin{myexample}[Enzyme kinetics] \label{example_1}

We consider the simple enzymatic reaction network
\begin{equation*}\label{michelis_menten}
\ch{ E + S <=>[$k\sb{1}$][$k\sb{-1}$] ES ->[$k\sb{2}$] E + P }.
\end{equation*}
In this CRN, the set of chemical species is $\mathcal{S}$ = \ch{(E, S, ES, P)}. The enzyme \ch{E}, binding the substrate \ch{S}, forms a complex \ch{ES}, which releases the product \ch{P} and the original enzyme \ch{E}.
According to \emph{mass-action kinetics}, the rate of each reaction is directly proportional to the concentrations of its reactants, via a coefficient marked next to each arrow ($k_1$,$k_{-1}$ and $k_2$, respectively). In our framework, this system is represented with a reversible reaction $\mathcal{R}_1$ and an irreversible one $\mathcal{R}_2$, with rate vector
\[
R(S) = \begin{bmatrix}
k_1 [E][S] - k_{-1}[ES]\\
k_2[ES]
\end{bmatrix}.
\]
Here, we have used the symbol $[E]$ to denote the concentration of \ch{E}, and so on.

The  differential equations describing the behavior of the CRN are
\[
\begin{cases}
&\frac{d[E]}{dt}=-k_{1}[E][S] + k_{-1}[ES] + k_2[ES],\\
&\frac{d[S]}{dt}=-k_{1}[E][S]+ k_{-1}[ES],\\
&\frac{d[ES]}{dt}=+k_{1}[E][S] - k_{-1}[ES] - k_2[ES],\\
&\frac{d[P]}{dt}= +k_{2}[ES].\\
\end{cases}
\]
\end{myexample}

As we can notice, a CRN is usually described by a set of non-linear differential equations, which makes difficult the analysis of the dynamics of the system. Indeed, different initial concentrations of the species involved in the reactions can affect the internal and external behavior of the network. Besides, the exact values of parameters are often unknown. Then, in order to investigate the behaviour of the system, we need to study it by performing many simulations under  all possible combinations of chemical species concentrations. 

To reduce the computational effort, one of the possible approaches is to  study  the qualitative behaviour of CRNs, without making assumptions on the structure of the dynamical equations involved. In this context, establishing some  kind of monotonicity property   can help to answer questions concerning the network asymptotic dynamics, such as which are  the functionalities of specific chemical pathways or how  parameter variations influence the network. Indeed, monotonicity describes  the capacity of a system to respond in a natural way  to perturbation on its components. 

According to Angeli et al.~\citep{angeli2006structural, angeli2010graph} a system is monotonic if its forward flow preserves some order defined on the state space. The system dynamics is expressed in terms of \textit{reaction coordinates}, for which all the reaction processes are taken into account as a unique flux. As a consequence, with this approach, 
 while it is possible to study  how internal or external perturbations influence the fluxes generated by the reactions,
it is not possible to address (and therefore understand) how  different concentrations of the chemical species involved in the network influence the overall behaviour. 
The  main advantage of Angeli's approach is that  the authors provide efficiently verifiable
sufficient conditions to establish if a system is globally monotonic, based on a graphical representations of reactions. In particular, they investigate the dynamics of the system using a particular kind of graph, the so called \textit{R-graph}.
Their sufficient condition for global monotonicity is the following: the network is globally monotonic if each closed path of the R-graph contains an even number of negative edges (\textit{positive loop property}). In the next section, we shall see a formal definition of this graph and comment on this condition.

\section{Results}\label{sec:Results}
\subsection*{Input-Output monotonicity definition}

Global monotonicity, proposed in \citep{angeli2006structural}, is a very strong property, since it is based on an unique ordering on the whole reaction network. Unfortunately, such a strong property does not hold on most realistic chemical reaction networks.

For this reason, our goal is to assess  monotonicity properties between species concentrations that would allow us to  
infer the behaviour of the network under different initial conditions and to define  sufficient conditions for chemical reactions networks that are easy to test and guarantee that such monotonicity properties hold.

The monotonicity property we are interested in can be summarised as follows. Given two species, that we call \textit{input} and \textit{output} species of the network, we say that there is a monotonicity relation if the concentration of output species at any time either increases or decreases due to an increase in the initial concentration of the input species. If established, our 
monotonicity property would allow us to  substantially reduce the number of simulations required to study the system. Indeed, if two species are in a monotonicity relation and we are interested in studying the dynamics of the output when the input varies, we can avoid to simulate the chemical reaction network for all possible values of the initial concentrations of the input species.
To formally define the previous intuition,  we give a new definition of monotonicity, namely the \textit{Input-Output monotonicity}, and then,
 following the approach  in  \citep{angeli2010graph}, we give sufficient conditions that  guarantee the monotonicity relation of the input and  output species.


The following two definitions describe the concept of monotonicity we are interested in: it describes whether the output species reacts in a monotonic way to the increase of the input concentration. 

We consider two initial states $S^0, \overline{S^0}$ such that $\overline{S^0_I} > S^0_I $ for one particular species $I$ (the \emph{input species}), and $S^0_k = \overline{S^0_k}$ for all other species $k\neq I$.
With  $S_i(t)$ we indicate  the solution of the ODEs for the species $S_i$ with initial value $S(0) = S^0$, and with $\overline{S_i}(t)$ the solution with initial value $\overline{S}(0) = \overline{S^0}$.

\noindent \begin{Definition}[Positive Input-Output Monotonicity]
Given a set of reactions $\mathcal{R}$, species $\mathcal{S}_O \neq \mathcal{S}_I$ is \emph{positively monotonic} with respect to $\mathcal{S}_I$ in $\mathcal{R}$ if, for any two initial states $S^0, \overline{S^0}$ as above, $\overline{S_O}(t) \geq S_O(t)$, for every time $t\in \R_{\geq 0}$.
\end{Definition}

\noindent \begin{Definition}[Negative Input-Output Monotonicity]
Given a set of reactions $\mathcal{R}$, species $\mathcal{S}_O \neq \mathcal{S}_I$ is \emph{negatively monotonic} with respect to $\mathcal{S}_I$ in $\mathcal{R}$ if, for any two initial states $S^0, \overline{S^0}$ as above, $\overline{S_O}(t) \leq S_O(t)$, for every time $t\in \R_{\geq 0}$.
\end{Definition}

\subsection*{Directed SR-graph and R-graph}

We define two graphs associated to a reaction network. The directed SR-graph models the interplay between species and reactions.
\begin{Definition}[Directed SR-graph] \label{directed_SR_Graph}
    Given a finite set of reactions $\mathcal{R}$ over a set of species $\mathcal{S}$, the \emph{directed SR-graph} is the directed graph $\langle \mathcal{S} \cup \mathcal{R}, E \rangle$, where $E \subseteq (\mathcal{S} \cup \mathcal{R})\times (\mathcal{S} \cup \mathcal{R})$ is defined as follows.
    \begin{itemize}
        \item If $\Gamma_{jk} \neq 0$, i.e., $\mathcal{R}_k$ affects the concentration of $\mathcal{S}_j$, then we include the edge $(\mathcal{R}_k, \mathcal{S}_j)$ in $E$. 
        \item If $DR_{ij} \neq 0$, i.e., $\mathcal{S}_j$ affects the speed of $\mathcal{R}_i$, then we include the edge $(\mathcal{S}_j, \mathcal{R}_i)$ in $E$.
    \end{itemize}
    No other edges exist in $E$ (in particular, there no edges in $\mathcal{S}\times \mathcal{S}$ nor in $\mathcal{R} \times \mathcal{R}$.
\end{Definition}
In particular, for each pair $(\mathcal{R}_i, \mathcal{S}_j)$ so that $\mathcal{S}_j$ is involved in $\mathcal{R}_i$, one of these three cases applies.
\begin{itemize}
    \item[C1] if $\mathcal{S}_j$ is a promoter for $\mathcal{R}_i$ that does not get consumed, i.e., $DR_{ij}\neq 0$ but $\Gamma_{ji}=0$, then the edge $\mathcal{S}_j \to \mathcal{R}_i$ exists only in this direction;
    \item[C2] if $\mathcal{S}_j$ is produced or consumed by $\mathcal{R}_i$ but does not affect its rate (for instance the product of an irreversible reaction), i.e., $DR_{ij} = 0$ but $\Gamma_{ji} \neq 0$, then the edge $\mathcal{R}_i \to \mathcal{S}_j$ exists only in this direction;
    \item[C3] in all other cases in which $\mathcal{S}_j$ is involved $\mathcal{R}_i$, the edge $\mathcal{R}_i \leftrightarrow \mathcal{S}_j$ is bidirectional.
\end{itemize}

\begin{Definition}[Consistent labeling] \label{consistent_labeling_definition} A \emph{consistent labeling} of the set of reactions $\mathcal{R}$ is a map $\sigma: \mathcal{R} \to \{+1, -1\}$ such that
\begin{equation} \label{monocond}
    \sigma(i) DR_{ij} \Gamma_{jk} \sigma(k) \geq 0, \quad \forall i\neq k, \, \forall j.
\end{equation}

\end{Definition}

The second graph that we define is a signed but undirected graph that represents the constraints for the existence of a consistent labeling.
\begin{Definition}[R-graph]\label{R-graph_def}
Given a finite set of reactions $\mathcal{R}$ over a set of species $\mathcal{S}$, the R-graph of $\mathcal{R}$ is the signed graph $\langle \mathcal{R}, E_{+}, E_{-} \rangle$, where $E_{+} \subseteq (\mathcal{R} \times \mathcal{R}) $ and  $E_{-} \subseteq (\mathcal{R} \times \mathcal{R})$ are defined as follows:
\begin{itemize}
    \item $(\mathcal{R}_i, \mathcal{R}_k) \in E_{+} $ if $i\neq k$ and there is a species $\mathcal{S}_j$ such that $DR_{ij}\Gamma_{jk}>0$.
     \item $(\mathcal{R}_i, \mathcal{R}_k) \in E_{-} $ if $i\neq k$ and there is a species $\mathcal{S}_j$ such that $DR_{ij}\Gamma_{jk}<0$.
\end{itemize}
\end{Definition}
Note that an edge in $E_+ \cup E_-$ exists if and only if there is a path $\mathcal{R}_i \leftarrow \mathcal{S}_j \leftarrow \mathcal{R}_k$ in the directed SR-graph, so the R-graph is the graph obtained from the SR-graph by removing the vertices corresponding to species and connecting reactions directly.

In biological terms, edges of the R-graph describe the relationship between reactions behavior. If the reactions cooperate, helping each other, the reaction nodes are linked by a positive edge. This occurs when the chemical species produced by a reaction is the reactant of an another one. Otherwise, if the reactions compete, for example when they share the same reactants, the reaction nodes are linked by a negative edge. 

Note tht these definitions are very similar, but subtly different, from those in~\citep{angeli2006structural,angeli2010graph}: indeed, in those papers the case of enzymes (C1) is excluded \emph{a priori}, and the SR-graph is constructed in an undirected version based only on $\alpha_{ij}$ and $\beta_{ij}$, neglecting the difference between cases C2 and C3.

\begin{myexample} \label{ex:difference}
Consider the network
\begin{equation} \label{difference}
    \mathcal{R}_1: A \to C, \quad \mathcal{R}_2: B \to C, \quad \mathcal{R}_3: C \to D,
\end{equation}
where all reactions are irreversible. According to our definition, there is no edge between $\mathcal{R}_1$ and $\mathcal{R}_2$ in the R-graph. Under the definition in~\citep{angeli2006structural,angeli2010graph}, there is a negative edge between them, instead.
\end{myexample}

The following graph-theoretical characterization appears in~\citep{angeli2006structural}.
\begin{Lemma}
    Given a finite set of reactions $\mathcal{R}$ over a set of species $\mathcal{S}$, a consistent labeling of $\mathcal{R}$ exists if and only if every loop in the R-graph includes an even number of edges in $E_-$ (the \emph{positive-loop property}).
\end{Lemma}
Indeed, to construct a consistent labeling, we can assign any value to $\sigma(\mathcal{R}_1)$ and then traverse the graph assigning signs so each encountered vertex so that~\eqref{monocond} holds; if the graph has the positive-loop property we will never encounter a contradiction.

In addition, we can find a necessary condition for this labeling to exist.
\begin{Lemma}[Rule of 2] \label{lem:rule2}
For each species $\mathcal{S}_j$ in a reaction network, determine a number $n(\mathcal{S}_j)$ as follows. Consider all reactions in which $\mathcal{S}_j$ is involved, and determine which of the cases C1, C2, C3 above holds.
\begin{itemize}
    \item Take the number of reactions in case C3.
    \item Add 1 if there are one or more reactions that fall in case C1.
    \item Add 1 if there are one or more reactions that fall in case C2.
\end{itemize}
Call this number $n(\mathcal{S}_j)$
If, for some species $\mathcal{S}_j$, $n(\mathcal{S}_j) > 2$, then $\mathcal{R}$ does \emph{not} admit a consistent labeling.
\end{Lemma}
\begin{proof}
 Suppose by contradiction that a consistent labeling exists, but that $n(\mathcal{S}_j) > 2$ for a certain species $\mathcal{S}_j$. Then, we can find reactions $i,k,\ell$ so that $\mathcal{R}_i$ falls in case C1 or C3, $\mathcal{R}_k$ falls in case C2 or C3, and $\mathcal{R}_\ell$ falls in case C3. Thanks to the definition of the three cases and to~\eqref{monocond},
 \begin{align*}
     \sigma(i) DR_{ij} \Gamma_{j\ell} \sigma(\ell) &> 0,\\
     \sigma(i) DR_{ij} \Gamma_{jk} \sigma(k) &> 0,\\
     \sigma(\ell) DR_{\ell j} \Gamma_{j\ell} \sigma(\ell) &> 0.
 \end{align*}
Taking the product of these three inequalities, one obtains $ DR_{\ell j} \Gamma_{j\ell} > 0$, which contradicts our assumption~\eqref{assumption}.
\end{proof}

\begin{myexample*}[\ref{ex:difference}, continued]
In the network~\eqref{difference}, the only species involved in more than one reaction is C. We have $n(C)=2$, as it is involved in $\mathcal{R}_3$ under case C3 (1), and in $\mathcal{R}_1,\mathcal{R}_2$ under case C2 (+1). Thus Lemma~\ref{lem:rule2} does \emph{not} prohibit the existence of a consistent labeling. Indeed, one can verify that $\sigma(\mathcal{R}_1)=\sigma(\mathcal{R}_2)=\sigma(\mathcal{R}_3)=1$ is a consistent labeling. The same network~\eqref{difference}, does not admit a consistent labeling with the definitions in~\citep{angeli2006structural,angeli2010graph}, instead, as Condition~2 in \citep[Proposition~1]{angeli2006structural} fails with their definition.
\end{myexample*}

\subsection*{Monotonicity in reaction coordinates}

Let $x(t) = \int_{0}^t R(S(\tau)) d\tau$ be the vector such that $x_i$ is the extent of the $i$th reaction. The vector $x(t)$ solves, with initial condition $x(0)=0$, the system of differential equations
\begin{equation} \label{reactionsystem}
    \frac{d}{dt}x(t) = R(S^0 + \Gamma x(t)),
\end{equation}
where $S^0$ is the vector of initial concentrations.

The following result appears in~\citep{angeli2006structural}, and continues to hold with our definition of R-graph (as proved in Section~\ref{sec:Methods}).
\begin{Theorem} \label{thm:angeli}
    If the R-graph has the positive-loop property, then the system~\eqref{reactionsystem} is \emph{orthant-cooperative} (or \emph{orthant-monotone}), i.e., for a certain diagonal matrix $\Sigma$, given two initial conditions $x^0$ and $\overline{x^0}$ such that $\Sigma(x^0 - \overline{x^0}) \geq 0$, then the solutions $x(t),\overline{x}(t)$ of~\eqref{reactionsystem} with those initial conditions satisfy $\Sigma(x(t) - \overline{x}(t)) \geq 0$ for each time $t\geq 0$.
\end{Theorem}
An explicit choice for this matrix $\Sigma$ is obtained by taking $\Sigma_{ii} = \sigma(i)$, where $\sigma$ is a consistent labeling of $\mathcal{R}$ (which exists thanks to the positive-loop property).

Note that the physical meaning of Theorem~\ref{thm:angeli} is not apparent, since $x^0=0$ (i.e., empty history of reactions occurred before the initial time) is the only initial condition that has a direct relevance to the application to CRNs. Rather, this theorem is used in~\citep{angeli2006structural,angeli2010graph} to obtain statements about the asymptotic behavior of the reaction network.

\subsection*{Handling reversible reactions}

As in \citep{angeli2006structural}, in the case of reversible reactions, we construct the graph by considering only one of the two possible orientations. The choice of reaction orientation does not influence Angeli's monotonicity result that we will exploit in the proof of our theorems. Moreover, even if this choice changes the way we process the R-graph, as we will discuss later, it does not influence the final result of our method.


Note that, as we already pointed out, the choice of the orientation of reversible reactions does not influence the existence of a consistent labeling.

\begin{Lemma} Given a set of reactions $\mathcal{R}$ and a reaction $\mathcal{R}_i \in \mathcal{R}$, let $\mathcal{R}^{rev}_i$ be the reaction obtained from $\mathcal{R}_i$ by swapping reactants with products. There exists a consistent labeling for the R-graph of $\mathcal{R}$ if and only if there exists a consistent labeling for the R-graph of $\mathcal{R}' = \mathcal{R}\setminus \{\mathcal{R}_i\} \cup \{\mathcal{R}^{rev}_i\}$.
\end{Lemma}

\begin{proof}
Let $\langle \mathcal{R}, E_{+}, E_{-} \rangle$ be the R-graph of $\mathcal{R}$. The R-graph of $\mathcal{R}'$ can be obtained from the previous one as follows: $\langle \mathcal{R}', E'_{+}, E'_{-} \rangle$ where $E'_{+} = E_{+} \setminus \{ (\mathcal{R}_i, \mathcal{R}_j) \in E_{+} \} \cup \{ (\mathcal{R}^{rev}_i, \mathcal{R}_j) \mid (\mathcal{R}_i, \mathcal{R}_j) \in E_{-} \}$ and $E'_{-} = E_{-} \setminus \{ (\mathcal{R}_i, \mathcal{R}_j) \in E_{-} \} \cup \{ (\mathcal{R}^{rev}_i, \mathcal{R}_j) \mid (\mathcal{R}_i, \mathcal{R}_j) \in E_{+} \}$. Hence, the R-graph of $\mathcal{R}'$ is the same as the one of $\mathcal{R}$, but in which edges that were connected to $R_i$ are now connected to $R^{rev}_i$ with opposite sign. Given a consistent labeling for the R-graph of $\mathcal{R}$ it is possible to derive a consistent labeling for $\mathcal{R}'$ by simply assigning to $\mathcal{R}^{rev}_i$ an opposite label with respect to $\mathcal{R}$, and vice-versa.
\end{proof}

\subsection*{Sufficient condition for input-output monotonicity }

Our main result is the following, proved in Section~\ref{sec:Methods}.

\begin{Theorem} \label{THEOREM}
Let a set of chemical reactions $\mathcal{R}$ be given, with two distinct species $\mathcal{S}_I$ and $\mathcal{S}_O$ designated as the input and the output species.

Augment the network by adding two dummy reactions: an irreversible reaction with output $\mathcal{S}_I$
\[
\mathcal{R}_{IN} : \emptyset \to \mathcal{S}_I,
\]
and a reaction which has $\mathcal{S}_{O}$ as promoter
\[
\mathcal{R}_{OUT} : \mathcal{S}_{O} \to \mathcal{S}_{O}.
\]
If $\mathcal{R} \cup \{\mathcal{R}_{IN}, \mathcal{R}_{OUT}\}$ has a consistent labeling (i.e., if its R-graph has the positive-loop property), then $\mathcal{S}_O$ is monotonic with respect to $\mathcal{S}_I$. In particular, it is positively monotonic if $\sigma(IN)\sigma(OUT)=+1$, and negatively monotonic if $\sigma(IN)\sigma(OUT)=-1$.
\end{Theorem}



To show how Theorem \ref{THEOREM} works, consider again our running Example \ref{example_1}, where we want to study the monotonicity between the species \ch{S} and the species \ch{P}, which we consider as the input and the output of the CRN.

We build the stoichiometric matrix of the network, considering only one direction for the  reversible reaction $\mathcal{R}_1$
\[
\Gamma = \bordermatrix{
~ & \mathcal{R}_1 & \mathcal{R}_2 \cr
E & -1 & +1\cr
S & -1 & 0\cr
ES & +1 & -1\cr
P & 0 & +1\cr
},
\]
and the matrix $DR$
\[
DR = \bordermatrix{
~ & E & S & ES & P \cr
\mathcal{R}_1 & +k_1[S] & +k_1[E] & -k_1[ES] & 0 \cr
\mathcal{R}_2 & 0 & 0 & +k_2 & 0\cr
}.
\]

\begin{figure}[t]
\begin{center}
 \centerline{ \includegraphics[width=0.55\linewidth]{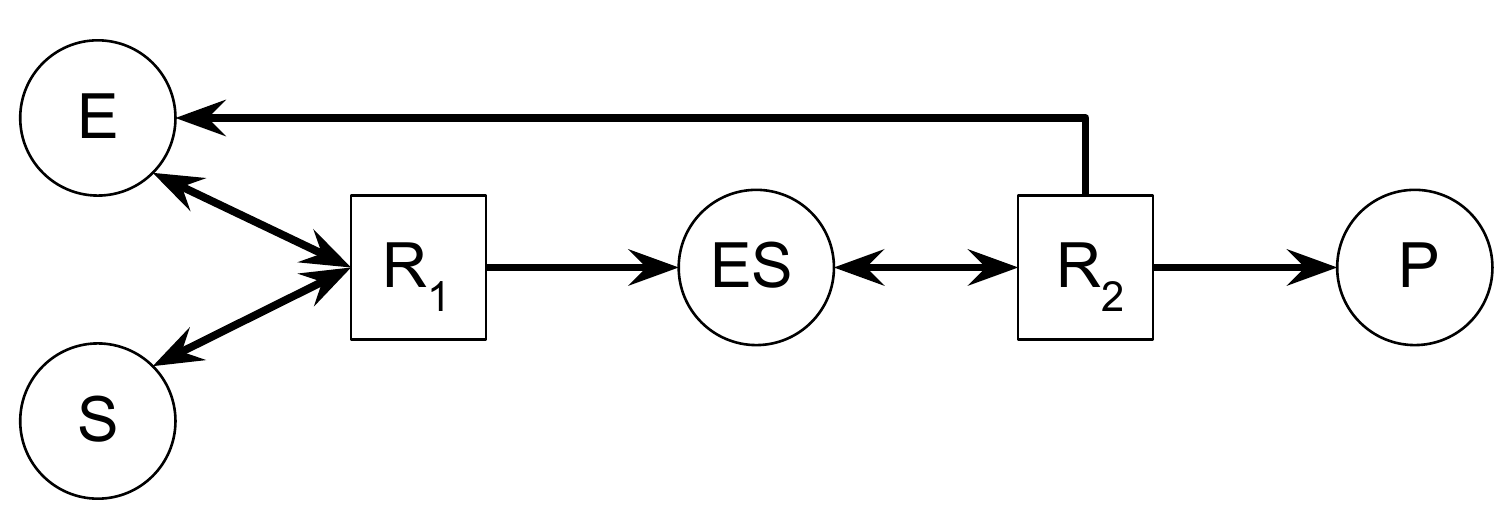}}
\end{center}
\caption{Directed SR-graph of Example \ref{example_1}, representing Michaelis-Menten kinetics.}
\label{Figure_1}
\end{figure}

\begin{figure}[t]
\begin{center}
 \centerline{ \includegraphics[width=0.2\linewidth]{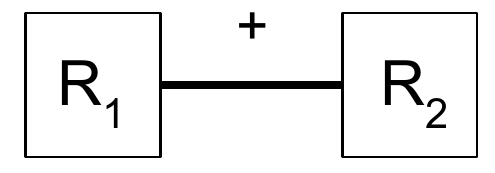}}
\end{center}
\caption{The R-graph of Example \ref{example_1}, representing Michaelis-Menten kinetics.}
\label{Figure_R-graph}
\end{figure}

Then, following Definition \ref{directed_SR_Graph}, we build the directed SR-Graph, represented in Figure \ref{Figure_1}, from which we derive the R-graph, which has one positive edge since we obtain 
$DR_{ij}\Gamma_{jk} > 0$ for each species $j$. We represent the R-graph in Figure \ref{Figure_R-graph}.

Then, we augment the network adding the two following dummy reactions: 
\begin{align*}
\begin{split}
&\ch{0 -> S} \\
&\ch{P->P}, \\
\end{split}
\end{align*}
and we verify trivially that the new network has a consistent labeling, which means that output is monotonic with respect to the input. Moreover, since $\sigma(IN)\sigma(OUT)= +1$ we can affirm that the output of this CRN, the chemical species \ch{P}, is \textit{positively} monotonic with respect to the species \ch{S}, which means that if we increase the initial concentration of the input \ch{S}, then the concentration of the output \ch{P} increases at any time. 

\begin{figure}[t]
 \centerline{\includegraphics[width=0.6\linewidth]{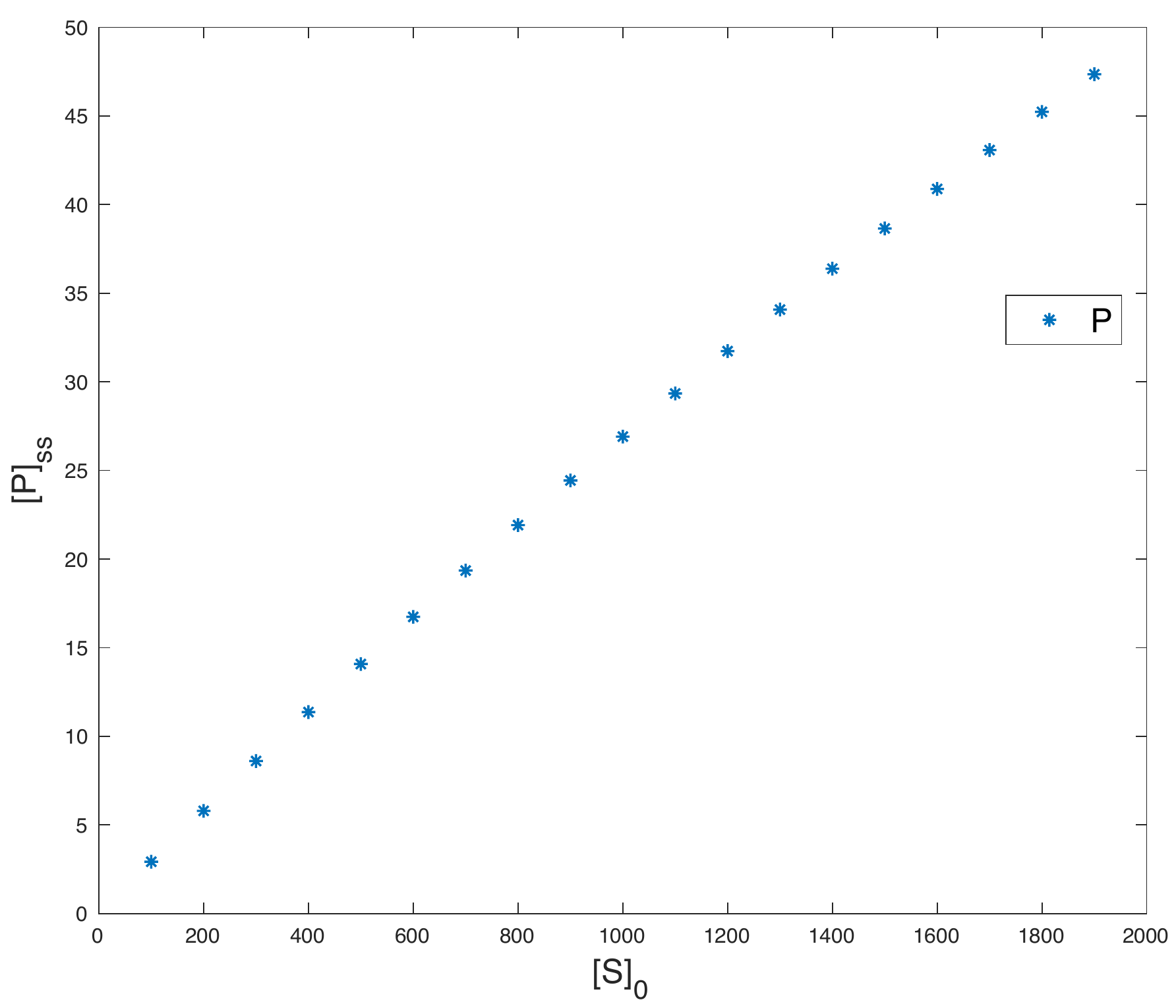}}
  \caption{Simulation results of Example \ref{example_1}, representing Michaelis-Menten kinetics. To show how the concentration of the species P is positively monotonic with respect to the species S, we plot on the horizontal axis the initial concentration of S, in a range $[100, 2000]$ and on the vertical axis the concentration of P at the steady state.}
      \label{Figure_2}
      \end{figure}

As expected, this result is  confirmed by simulations. Once we have fixed the chemical rates $(k_1=0.1, k_2=1000, k_3=0.3)$, we
vary the initial concentration of the input \ch{S} in the range $[100,2000]$. In Figure \ref{Figure_2}, we show that by increasing the initial concentration of \ch{S} (denoted $[S]_O$), the concentration of \ch{P} at the steady state (denoted $[P]_{ss}$) increases as well. This actually does not hold only for the steady state: the concentration of \ch{P} is increased at any time point. 


\section{Case study}\label{sec:Case}
We now  apply the result stated in  Theorem \ref{THEOREM} to the more complex network of ERK signalling pathway.  

\subsection*{Signalling pathway example.} A signalling pathway usually consists of enzymatic cascades, having a starting species that triggers the other connected processes. An initial stimulus, perceived by a \textit{transductor} (a sort of the first messenger), activates the cascade amplifying the signal for the next enzymatic reaction. Many biochemical processes are associated with signalling pathways as protein activation, repression, and expression of genes: anomalies in these processes could give rise to diseases like cancer, diabetes, and others. 

One of the most important examples of such processes is the ERK pathway, which is involved in growth, survival, proliferation, and differentiation of cells. We consider the mathematical model of the ERK pathway implemented by Schilling et al. \citep{schilling2009theoretical} and available to the public on the BioModels Database (BIOMD0000000270). 
It consists of many fast phosphorylation reactions, which spread the signal along the enzymatic cascade. For simplicity, let us focus on a particular portion of the entire pathway, which we will denote as ERK$*$. We indicate the species and the kinetics rates as originally denoted in the model in \citep{schilling2009theoretical}. For simplicity, we refer to the reaction using the notation $R_i$, where $i$ is the kinetics rate index. The reactions involved are the following: 

\begin{align}
\begin{split}
&\ch{Raf <=>[$k\sb{18}$][$k\sb{19}$] PRaf} \\
&\ch{Mek1<=>[$k\sb{21}$[PRaf]][$k\sb{27}$]PMek1} \\
&\ch{PMek1<=>[$k\sb{23}$][$k\sb{25}$]PPMek1}, \\
\label{erk_reaction1}
\end{split}
\end{align}
in Table \ref{Table_example2} we reported the coefficient rates and the initial conditions of ERK$*$ system.

\begin{deluxetable*}{cc}
\tablenum{1}
\tablecaption{The initial concentrations and the rates of ERK* system.\label{Table_example2}}
\tablewidth{0pt}
\tablehead{
\colhead{Initial concentrations} & \colhead{Rates}}
\decimalcolnumbers
\startdata
Raf = 10             & $k_{18}= 0.1445$\\  
Praf = 0             & $k_{19}= 0.37 $   \\ 
Mek1= 1              & $k_{21}= 0.02$    \\ 
PMek1 = 0            & $k_{23}= 667.957$   \\
PPMek1 = 0           & $k_{25}= 0.13 $    \\
                     &  $ k_{27}= 0.07$  \\ 
\enddata

\end{deluxetable*}


The species PRaf, in the reaction $R_{21}$, is involved as \textit{catalyst promoter}, which means that its concentration positively influences the production of the species PMek1. In our framework, the rate vector for this network is
\[
R(S) = \begin{bmatrix}
k_{18}[Raf] - k_{19}[PRaf]\\
k_{21}[Mek1][PRaf] - k_{27}[PMek1]\\
k_{23}[PMek1] - k_{25}[PPMek1]
\end{bmatrix}.
\]

We can apply Theorem \ref{THEOREM} on the network, considering Raf and PPMek1, respectively, as the input and the output. We augment the network adding the two following dummy reactions: 
\begin{align*}
\begin{split}
&\ch{0 -> Raf} \\
&\ch{PPMek1->PPMek1}. \\
\end{split}
\end{align*}

\begin{figure}[t]
 \centerline{\includegraphics[width=0.55\linewidth]{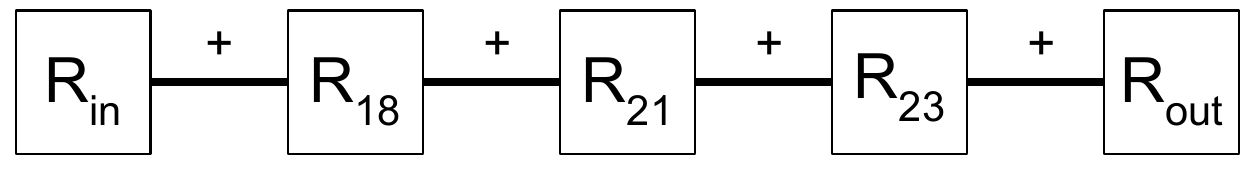}}
  \caption{R-graph representation of the network \ref{erk_reaction1}, augmented by the two dummy reactions. As we can notice it has a consistent labeling.}
      \label{Figure_6s}
      \end{figure}

\begin{figure}[h!]
 \centerline{\includegraphics[width=0.6\linewidth]{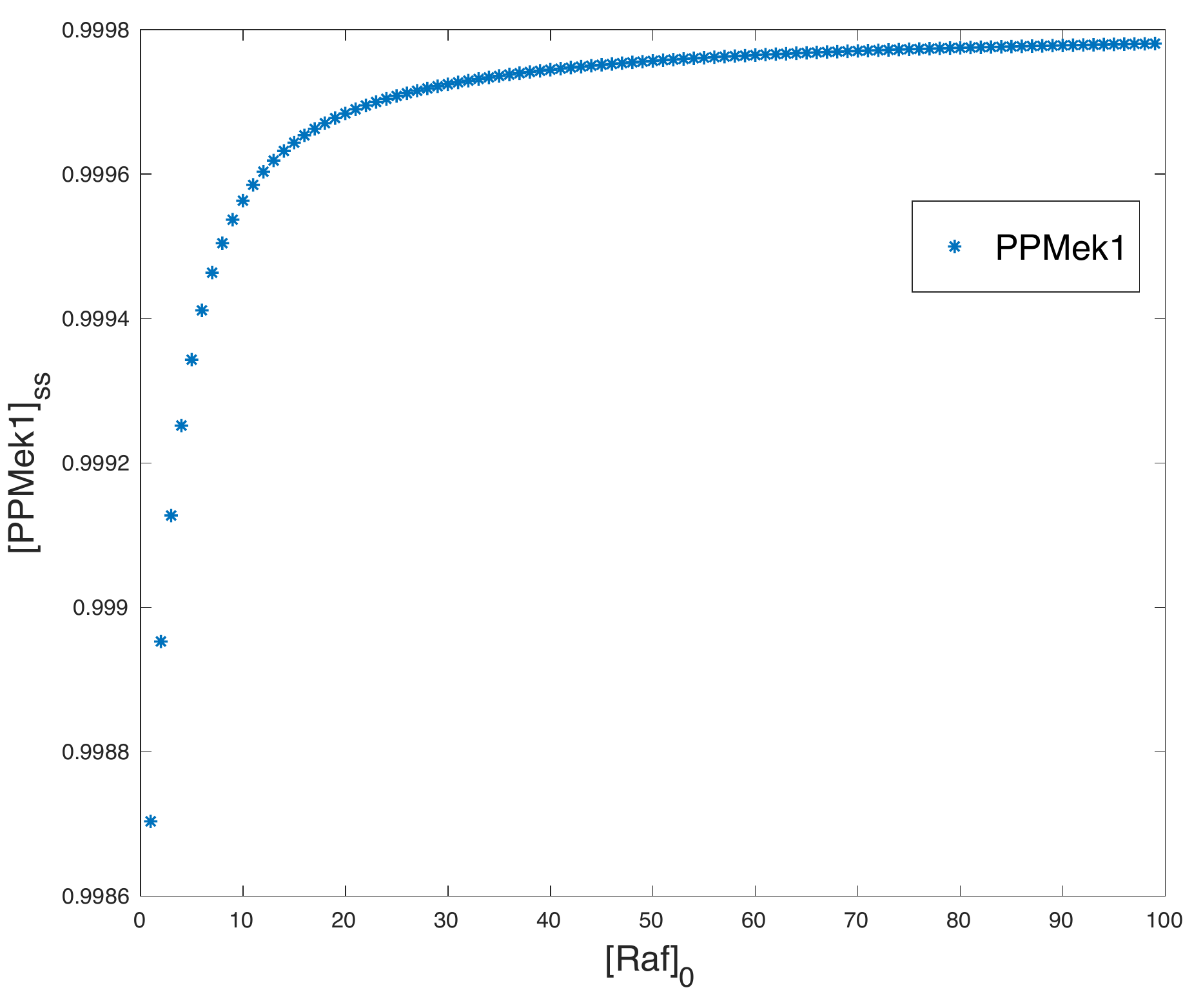}}
  \caption{Simulation results of CRN \ref{erk_reaction1}, representing ERK signalling pathway. To show how the concentration of the species PPMek1 is positively monotonic with respect to the species Raf, we plot on the horizontal axis the initial concentration of Raf, in a range $[1, 100]$ and on the vertical axis the concentration of PPMek1 at the steady state.}
      \label{Figure_6}
      \end{figure}

We build the R-graph and we can easily verify that it has a consistent labeling, as we show in Figure \ref{Figure_6s}. Moreover, since $\sigma(IN)\sigma(OUT)= +1$, we can conclude that PPMek1 is positively monotonic with respect to Raf. This can be confirmed with simulations, as shown in Figure~\ref{Figure_6}. 

\section{Methods}\label{sec:Methods}
In this section, we illustrate the proofs of the two main results.

\begin{proof}[Proof of Theorem~\ref{thm:angeli}] This proof is basically the same one that appears in~\citep{angeli2006structural}, but adapted to our slightly different definitions.
 
If the R-graph has the positive-loop property, then there is a consistent labelling $\sigma$; let $\Sigma \in \mathbb{R}^{r\times r}$ be the diagonal matrix with $\Sigma_{ii} = \sigma(i)$.
 The Jacobian of~\eqref{reactionsystem} is $J = DR \cdot \Gamma$. The off-diagonal entries of $\Sigma J \Sigma$ are given by
\[
\Sigma_{ii}J_{ik}\Sigma_{kk} = \sum_j \sigma(i) DR_{ij} \Gamma_{jk} \sigma(k), \quad i \neq k,
\]
where all summands are non-negative by the definition of consistent labeling~\eqref{monocond}. The fact that $\Sigma J \Sigma$ has non-negative diagonal elements is a sufficient condition for the system to be orthant-monotone (see~\citep{smith95book}).
\end{proof}

\begin{proof}[Proof of Theorem~\ref{THEOREM}]

Let $e_I\in\mathbb{R}^s$ (resp. $e_O$) be the vector that has an entry $1$ in the corresponding to $\mathcal{S}_I$ (resp. $\mathcal{S}_O$) and all other entries equal to $0$.

The rate and stoichiometry matrix of the augmented network are
\begin{equation*}
    \widehat{DR} = 
    \kbordermatrix{ & \\
    1\dots r & DR\\
    IN & 0\\
    OUT & k e_{O}^T}, 
    \quad
    \widehat{\Gamma} = 
    \kbordermatrix{
    & 1 \dots r & IN & OUT\\
    & \Gamma & e_{I} & 0
    }
\end{equation*}
and its Jacobian is
\[
\widehat{J} = \widehat{DR} \cdot \widehat{\Gamma} = \kbordermatrix{
& 1 \dots r & IN & OUT\\
1 \dots r & J & DR e_{I} & 0\\
IN & 0  & 0 & 0\\
OUT & e_{O}^T \Gamma & 0 & 0
}.
\]
Let $\sigma$ be a consistent labeling for the augmented network, and assume without loss of generality (up to replacing $\sigma$ with $-\sigma$) that $\sigma(IN) = 1$.

Hence $\Sigma DR e_{I} \geq 0$, $\sigma(OUT) e_{O}^T \Gamma \Sigma \geq 0$, and $\Sigma J \Sigma $ has non-negative diagonal entries.

We now take two initial values $S^0$ and $\overline{S^0} = S^0 + \delta e_I$ with $\delta > 0$, and aim to prove that $\sigma(OUT)(\overline{S}_{O}(t) - S_{O}(t)) \geq 0$ for each $t \geq 0$; indeed, this is the input-output monotonicity statement that we need to prove.

Define the dynamical system
\begin{equation} \label{modreactionsystem}
\begin{cases}
\frac{d}{dt} y(t) = R(S^0 + z e_{I} + \Gamma x(t)),\\
\frac{d}{dt} z(t) = 0.
\end{cases}
\end{equation}

Let
\[
\check{J} = 
\kbordermatrix{
& y & z\\
y & J & DR e_{I}\\
z & 0 & 0
}, \quad 
\check{\Sigma} = \kbordermatrix{
& y & z\\
y & \Sigma & 0\\
z & 0 & 1
}.
\]
It is simple to verify that $\check{J}$ is the Jacobian of~\eqref{modreactionsystem}, and that $\check{\Sigma} \check{J} \check{\Sigma}$ has non-negative off-diagonal, hence the system~\eqref{modreactionsystem} is orthant-monotone by the same result in~\citep{smith95book} that we have used in the proof of Theorem~\ref{thm:angeli}.

Direct verification shows that the solution of this system with initial value $y(0)=0, z(0)=0$ is $y(t)=x(t), z(t)=0$, whereas the solution $\overline{y}(t),\overline{z}(t)$ with initial value $\overline{y}(0)=0, \overline{z}(0) = \delta$ is $\overline{y}(t) = \overline{x}(t), \overline{z}(t) = \delta$, where  the quantity $\overline{x}(t)$ is defined analogously to~\eqref{reactionsystem} but with initial value $\overline{S^0}$. 

We have
\[
\begin{bmatrix}
0\\ \delta
\end{bmatrix} = 
\check{\Sigma} \begin{bmatrix}
\overline{y}(0)\\
\overline{z}(0)
\end{bmatrix} \geq 
\check{\Sigma} \begin{bmatrix}
y(0)\\
z(0)
\end{bmatrix}
=
\begin{bmatrix}
0\\ 0
\end{bmatrix}.
\]
By the orthant-monotonicity of~\eqref{modreactionsystem}, this implies
\[
\check{\Sigma} \begin{bmatrix}
\overline{y}(t)\\
\overline{z}(t)
\end{bmatrix} \geq 
\check{\Sigma} \begin{bmatrix}
y(t)\\
z(t)
\end{bmatrix} \quad \text{for all $t \geq 0$},
\]
and this means in particular that $\Sigma \overline{x}(t) \geq \Sigma x(t)$.

The concentrations of the output species $S_{O}(t),\overline{S}_{O}(t)$ at time $t$ with the two initial conditions $S^0,\overline{S^0}$ are given by
\[
S_{O}(t) = S^0_{O} + e_{O}^T \cdot \Gamma x(t), \quad \overline{S}_{O}(t) = S^0_{O} + e_{O}^T \cdot \Gamma \overline{x}(t),
\]
respectively ($S^0_{O} = \overline{S^0}_{O}$ because $\mathcal{S}_O \neq \mathcal{S}_I$), hence
\[
\sigma(OUT)(\overline{S}_{O}(t) - S_{O}(t)) = \underbrace{\sigma(OUT) e_{O}^T \Gamma \Sigma}_{\geq 0} \underbrace{\Sigma (\overline{x}(t) - x(t))}_{\geq 0} \geq 0,
\]
which completes the proof.
\end{proof}

%
%

\section{Discussion and Conclusions}\label{sec:Disc_Concl}

In this paper, we proposed a new notion of monotonicity, namely the Input-Output monotonicity, for which two species, considered as input and output of the network, are monotonic if the variation of the initial concentration of the input implies a monotonic variation in the concentration of the output.
Moreover, we established a new sufficient condition based on the R-graph that guarantees that such monotonicity property holds. 
We showed how this new notion of monotonicity can have a great practical impact. Indeed, monotonicity assessment can drastically reduce the number of simulations necessary to study the dynamical behaviour of a chemical reaction network under uncertain initial conditions. This can be very useful in several different cases, for example when initial concentrations are not exactly  known  or when the focus is on studying the effects  of perturbations of initial concentrations (as in the case of robustness). The proposed method allows us to study complex chemical reaction networks, performing a preliminary analysis of the system dynamics, in order to investigate different biological properties, such as robustness or sensitivity to perturbations. 

We have shown the application of our approach to the small example of Michaelis-Menten kinetics and to the quite complex model of the ERK signaling pathway \citep{kwang2003mathematical}.  

In order to apply our sufficient condition on larger and more complex network, different approaches of model reduction can be applied. For instance, in \citep{kuken} the authors show how to remove particular nodes of the network preserving the steady state fluxes of the system. Moreover, for analysing large-scale biochemical models, we can also simplify the model using the common approach of the \textit{separation of timescale} \citep{ingalls2013mathematical}, which allows us to consider the reaction networks as divided into processes having different timescales leading to an approximate -- but accurate --  version of the original model. 

In addition, in \citep{bove2019}, the authors propose an innovative approach based on machine learning on graphs to predict whether a biological system is robust studying its topological features. As future work, we intend to apply a similar method to predict  the monotonicity of a system.


\section*{Author's contributions}
     \textbf{Conceptualization \& Formal analysis:} R. Gori, P. Milazzo, L. Nasti, F. Poloni. \textbf{Simulations \& Experiments:} L. Nasti. \textbf{Writing original draft:} R. Gori, P. Milazzo, L. Nasti, F. Poloni. \textbf{Writing - review \& editing:} R. Gori, P. Milazzo, L. Nasti, F. Poloni.\vspace*{-12pt}

\section*{Funding}
R.~Gori, P.~Milazzo, L.~Nasti are partially supported by the University of Pisa's project PRA\_2020\_26 ``Metodi informatici integrati per la biomedica''. F.~Poloni is partially supported by the University of Pisa's project PRA\_2020\_61 ``Analisi di reti complesse: dalla teoria alle applicazioni'' and by GNCS/INDAM (Istituto Nazionale di Alta Matematica).\vspace*{-12pt}

%
%

\bibliography{bibliography}{}
\bibliographystyle{aasjournal}

\end{document}